# Mechanical compression regulates tumor spheroid invasion into a 3D collagen matrix.

Mrinal Pandey[1], Young Joon Suh[1], Minha Kim[2], Hannah Jane Davis[2], Jeffrey E Segall[3], and Mingming Wu[1]*

[1] Department of Biological and Environmental Engineering, 306 Riley-Robb Hall, Cornell University, Ithaca, NY 14853
[2] Department of Biological Sciences, 216 Stimson Hall, Cornell University, Ithaca, NY 14853
[3] Anatomy and Structural Biology, Albert Einstein College of Medicine, 1300 Morris Park Avenue, Bronx, New York 10461

*E-mail: mw272@cornell.edu

**Abstract:**

Uncontrolled growth of tumor cells in confined spaces leads to the accumulation of compressive stress within the tumor. Although the effects of tension within 3D extracellular matrices on tumor growth and invasion are well established, the role of compression in tumor mechanics and invasion is largely unexplored. In this study, we modified a Transwell assay such that it provides constant compressive loads to spheroids embedded within a collagen matrix. We used microscopic imaging to follow the single cell dynamics of the cells within the spheroids, as well as invasion into the 3D extracellular matrices (EMCs). Our experimental results showed that malignant breast tumor (MDA-MB-231) and non-tumorigenic epithelial (MCF10A) spheroids responded differently to a constant compression. Cells within the malignant spheroids became more motile within the spheroids and invaded more into the ECM under compression; whereas cells within non-tumorigenic MCF10A spheroids became less motile within the spheroids and did not display apparent detachment from the spheroids under compression. These findings suggest that compression may play differential roles in healthy and pathogenic epithelial tissues and highlights the importance of tumor mechanics and invasion.

**Introduction**

Solid tumor stress is a key indicator of tumor physiology. Clinically, the feel, touch, and shape of a solid tumor are important diagnostic methods for malignancy of the tumor[1-3]. In advanced tumors, rapid tumor cell growth in a confined tissue environment often leads to compressive stresses within the tumor core [4, 5], increasing the interstitial fluid pressure[6, 7], and collapsing vascular vessels within the breast tumor [5, 8]. The tension around the tumor periphery is also important. Cell traction forces re-organize and align collagen matrices surrounding the tumor and invasion into interstitial space [9-11]. Taken together the hypoxic tumor core[12], heightened interstitial fluid pressure[13, 14] and aligned collagen matrices around the tumor[10], are closely associated with solid tumor stress and are known critical indicators of the malignant state of solid tumors.

The mechanical environment of tumor cells can critically regulate tumor invasion. Landmark work from the Weaver lab has shown that tension within the tumor microenvironment promotes tumor progression[3]. Recently, work from our labs and others has shown that breast tumor cells and the cancer associated fibroblasts align, stiffen and plastically deform the surrounding ECM, and the stiffened ECM in return promotes cell force generation and invasion [15, 16]. Although how tensional field within the TME regulates tumor progression is extensively investigated, there is limited work on how compressive stress influences tumor mechanics and invasion [2]. This is in part due to the lack of tools that can provide controlled compressive stress for tumors in a physiologically realistic setting and at the same time compatible with optical microscopy.

Studies of roles of compression on tumor cell migration have mostly focused on single cells. In a 2D setting where cells were plated on a substrate, compression can enhance the migration and invasion of breast[17, 18], pancreatic[19] and brain cancer cells[20]. There is limited work in a more physiologically realistic 3D setting where cells were embedded within 3D ECMs[21, 22]. In native states, tumors are compressed in a confined space typically surrounded by ECMs. Previous work in our labs has shown that tumor spheroids are more sensitive to biochemical cues compared to single cells[23].

To further the compression studies to a more physiologically realistic setting, we modified a 3D Transwell assay to investigate the effect of compression on normal and breast tumor spheroids embedded within collagen. The Transwell assay is compatible with single cell imaging within as well as outside the tumor spheroid. We find differential responses from malignant breast tumor spheroids (MDA-MB-231) and non-tumorigenic epithelial spheroids (MCF10A) when compressed, with MDA-MB-231 cells becoming more motile and invasive while MCF10A cells becoming less motile in response to compression.

**Results and Discussion**

**A modified Transwell assay to apply static mechanical compression on tumor spheroids compatible with microscopic imaging at single-cell resolution.**

A commercial Transwell assay was modified to apply well-defined mechanical compression to spheroids embedded within a 3D collagen gel. The key modification to the Transwell assay was the addition of a 1 mm thick PDMS membrane patterned with a microwell (4 mm diameter and 115 µm depth) and placed at the bottom of a 24-well plate (Fig. 1a). The depth of the PDMS well, $h$, is critical because it determines the amount of compressive strain that can be applied to the spheroid. The compressive strain was defined as *(d - h)/d*, where d is the diameter of the spheroid. For each experiment, PDMS membranes with the microwell were first fabricated using a standard soft lithography technique and adhered to the bottom of a 24-well plate (Fig. 1a,1b). Spheroid-embedded collagen was then introduced into each of the PDMS wells, and the collagen was allowed to polymerize. Subsequently, the spheroids were compressed by placing a Transwell insert with a metallic weight on top. The bottom of the insert was ensured to be in direct contact with the top of the PDMS membrane (Fig. 1b). We note that the bottom of the Transwell insert is made of a 10 µm thick, porous polyester PET membrane with average pore size of 8 µm and the diameter of the insert is 6.5 mm. The PET membrane allowed continuous nutrient and gas exchange between the 3D spheroid culture and the surroundings for the duration of the experiments (16 hours). In a typical experiment, we used four out of the 24 wells, and the spheroids were observed using an automatic translational stage on the microscope.

To create spheroids of a targeted size, a previously developed microwell array device was used (see Fig. S1 and also ref. [24]) The diameters of the spheroids were calculated using bright-field microscopic images, as shown in Fig. 1c. The size of the spheroids depended on the initial cell seeding density, total number of incubation days, and size of the microwell. Our hydrogel-based array microwell device provides a robust way to create spheroids of well-defined sizes. In this study, the average diameters in the horizontal plane of the generated MDA-MB-231 tumor spheroids were 197.27 ± 3 µm in the control (uncompressed) and

211.68 ± 6 µm in the compressed condition (Fig. 1d). Similarly, the average diameter of MCF10A spheroids was 209.53 ± 3 µm in the control and 224.78 ± 5 µm in the compressed condition (Fig. 1d). The average compressive strain was determined as the difference between the average spheroid diameter (197.27 µm) and the height of the PDMS well (115 µm) divided by the spheroid diameter, which was 41% for the MDA-MB-231 spheroids. Similarly, the average compressive strain for the MCF10A spheroids was 44%.

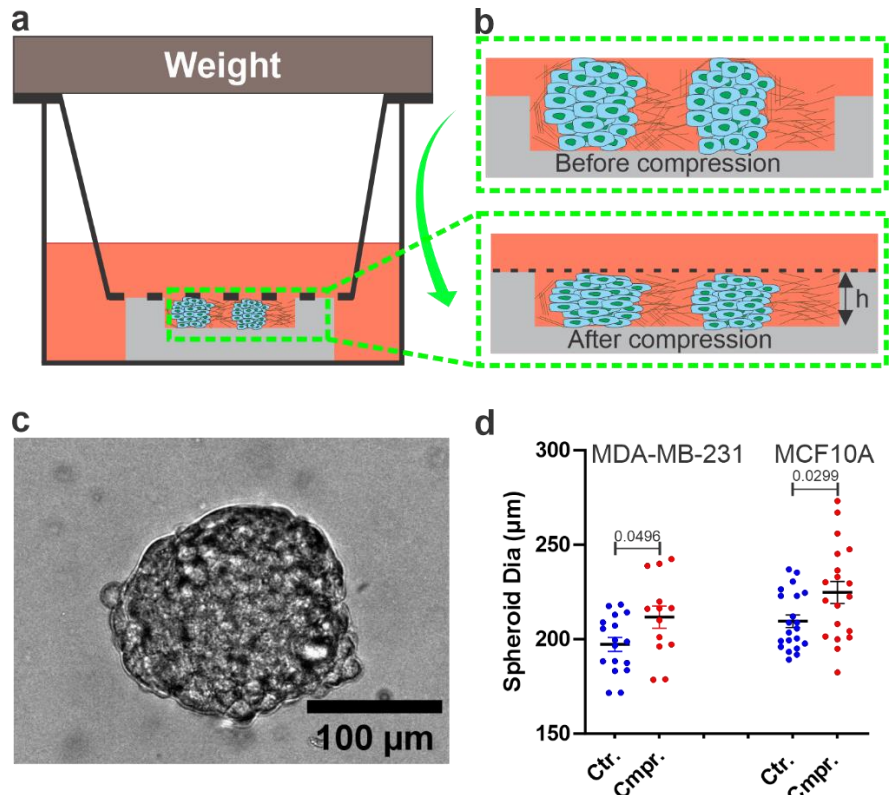

The unique capability of our modified Transwell assay is its compatibility with microscopic imaging, allowing us to follow single-cell dynamics within spheroids. In addition, it provides a straightforward method to apply compression to spheroids. Although only one compressive strain rate was used in our study, the compressive strain rate can be changed by using PDMS wells of different sizes. The disadvantage of our setup is that it does not allow for dynamic compression. To overcome this limitation, a microfluidic rheometer is currently being developed in our lab.

**Figure 1: In vitro model for mechanical compression of tumor spheroids: a)** Cross-sectional view of a modified Transwell assay for compressing tumor spheroids. Spheroid embedded collagen matrices were seeded within a PDMS well of defined height h, and a transwell insert with a fixed weight was placed on top of the PDMS well until the porous membrane of the insert and PDMS made contact. **b)** A detailed view of the cross-section of spheroid embedded collagen before and after compression. **c)** Brightfield image of an MDA-MB-231 spheroid surrounded by 3.5 mg/mL collagen, taken at t = 0, where t = 0 is the time imaging started. **d)** Scatter plot of MDA-MB-231 and MCF10A tumor spheroid diameters in the horizontal plane before and after compression. Each dot is data from one spheroid. The p values were obtained using a parametric Welch's t-test compared to the control group.

### Compression differentially regulates the motility of metastatic MDA-MB-231 and normal epithelial MCF10A cells within the spheroids.

Tumor mechanics has been a key indicator of tumor malignancy[25, 26]. To explore spheroid mechanics, we followed single-cell motility within spheroids.

*Following single cell motility within spheroids using fluorescently labeled cells.*

To study spheroid mechanics under compression, we investigated the motility of cells within tumor spheroids. To follow cell movement within the spheroid, we used a mix of 1:20 GFP labeled tumor cells to non GFP labeled tumor cells (Fig. 2a). 16-hour long time series of images of the fluorescent cells (see Fig. 2a and Movie S3 and S4) were obtained using an epi-fluorescence microscope. Each single-cell trajectory within the spheroid was tracked using the manual tracker in ImageJ, along with an in-house developed MATLAB code (Fig. 2b).

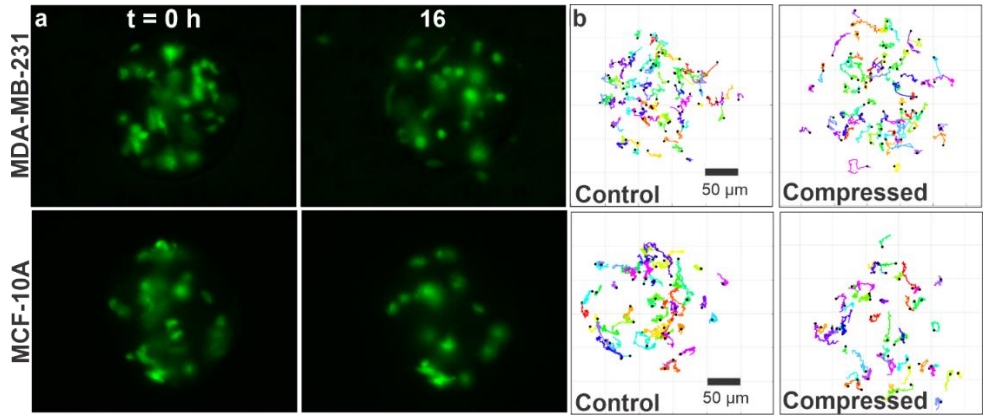

**Figure 2: Following single-cell dynamics within tumor spheroids. a)** GFP images of MDA-MB-231 and MCF10A tumor spheroids taken at t = 0 and t = 16 hours. A mixture of 1:20 fluorescent: non-fluorescent cells was used to make spheroids to facilitate single cell imaging within the spheroid. **b)** Trajectories of MDA-MB-231 and MCF10A cells within the tumor spheroids in control and compressed conditions. Each colored line is a cell trajectory, and cells were tracked before they invaded into the matrix. Each cell trajectory shown here is 10 hours long. A total of 80 cells were randomly tracked using 17 and 13 spheroids for control and compressed conditions for MDA-MB-231 and a total of 70 cells were tracked using 21 and 19 spheroids for control and compressed conditions for MCF10A spheroids, respectively.

*Compression enhanced motility of MDA-MB-231 cells but suppressed that of MCF-10A cells within spheroids.*

To quantify the motility of cells within the spheroids, we computed the cell motility parameters using the trajectories shown in Fig. 2b. Specifically, we obtained the speed, persistence length (plength), and mean squared displacement (MSDs) of the cells within the spheroid under compression and in the control (see Fig. 3). Within MDA-MB-231 spheroids, the cell speed was significantly enhanced by compression at an average speed of $0.117 \pm 0.003$ μm/min, in contrast to the control at $0.103 \pm 0.001$ μm/min (Fig. 3a,3b). There was no significant change in persistence length (Fig. 3c). One way to evaluate tumor spheroid mechanics is to use the diffusion coefficient, assuming that the cells are executing a random walk. To examine how cells diffuse within the spheroids, we computed the mean squared displacements (MSDs) and found that MSDs were greater in compressed in contrast to control in MDA-MB-231 spheroids (Fig. 3d). Using the first-order approximation for MSD, where MSD = 4Dt, we have the diffusion coefficient D = $0.047 \pm 0.0003$ μm$^2$/min for the compressed condition in contrast to D = $0.031 \pm 0.0005$ μm$^2$/min for the control. Similarly, we computed the motility parameters for MCF10A spheroids. Surprisingly, compression had the opposite effect on the MCF10A spheroids. The average speed of cells in compressed MCF10A spheroid condition was lower (0.102 ± 0.002 μm/min) than that of the control (0.154 ± 0.004 μm/min) (Fig. 3e,3f). The persistence length for MCF10A spheroids in the compressed state was 0.129 ± 0.009, in contrast to the control at 0.108 ± 0.009 (Fig. 3g), with no statistical significance observed. On examining the MSDs of the MCF10A cells within spheroids, we found that the MSD was significantly reduced upon compression, in contrast to the control (Fig. 3h). We computed the diffusion coefficient D = 0.030 ± 0.001 μm$^2$/min for the compressed condition, in contrast to D = 0.033 ± 0.0007 μm$^2$/min for the control. Interestingly, in MDA-MB-231 spheroids, we observed a time-dependent relationship; we found no major difference in speed of cells in the compressed and control conditions for first 8 hours and an increase in compressed condition was observed after this time point (see Fig. S2a). Whereas in MCF10A spheroids, the speed of cells in the compressed and control conditions were different from t=0 hour (see Fig. S2b).

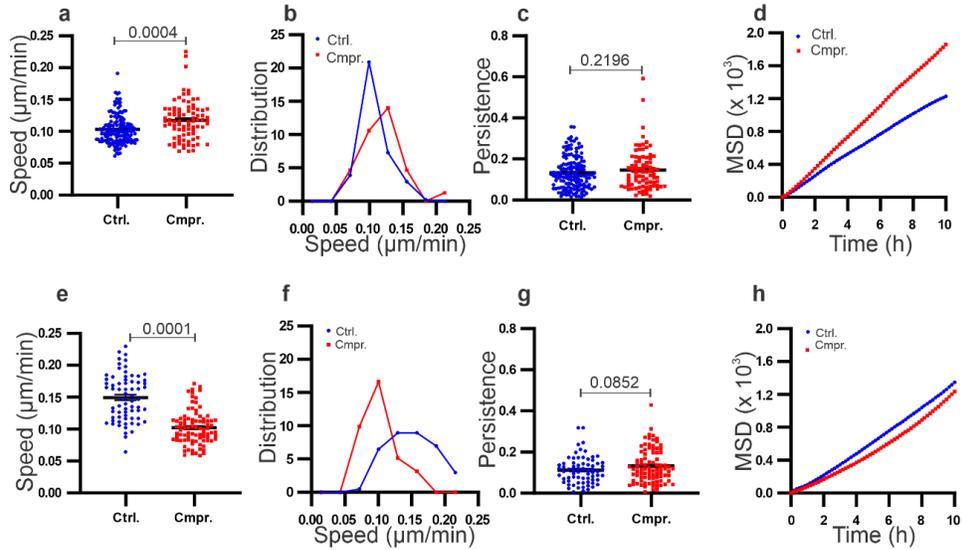

**Figure 3: Compression of tumor spheroids enhanced the motility of MDA-MB-231 cells but suppressed that of MCF10A cells within the spheroids. (a-d)** Cell migration speed (a), distribution of speed (b), persistence length (c), MSDs (d) of cells within MDA-MB-231 spheroids in control and compressed conditions. **(e-h)** Cell migration speed (e), distribution of speed (f), persistence length (g), MSDs (h) of cells within MCF10A spheroids in control and compressed conditions. The data here was computed using tracks shown in Fig 2b. The p values were obtained using Welch's parametric t-test compared to the control group.

Taken together, the speed and MSDs measurements suggest that cells within MDA-MB-231 tumor spheroids became more motile, whereas cells within MCF10A spheroids became slightly less motile upon compression. While the reason for this differential response requires further investigation, the work presented here is consistent with previous work carried out on the compression of confluent 2D cell sheets[18, 27]. In these studies, a wound healing assay was used to quantify the motility of cells. It was found that motility of malignant cells, including MDA-MB-231 and 4T1 cells, was enhanced when compressed, while motility of non-malignant MCF10A cells was suppressed upon compression. We also noted that in our compression study non-malignant MCF10A cells migrated faster than malignant MDA-MB-231 cells within the spheroids. Similar behavior has been reported previously, where MCF10A cells migrated faster than MDA-MB-231 cells when they were plated on a 2D substrate or embedded within a 3D collagen gel [17, 28].

**Compression differentially regulates invasion of MDA-MB-231 cells and MCF10A cells into the collagen matrix.**

Tumor cell invasion is an important step during cancer metastasis[29, 30]. Here, we asked whether compression regulates tumor invasion into 3D ECMs.

*Compression decreases the circularity of MDA-MB-231 spheroids but does not have observable effects on MCF10A spheroids.*

Time-lapse brightfield images of spheroids embedded in ECM were recorded with and without compression (Fig. 4a). The circularity parameter, defined as the ratio of the 4 * pi area to the perimeter squared, was used to quantify the spheroid shape. Note that a circularity value of 1 (maximum) indicates that the spheroid is perfectly circular, and a decreasing value from 1 reflects a deviation from circularity. The outlines of the spheroids were traced manually, as shown by the bright yellow lines in Fig. 4a, and the parameters were calculated using ImageJ. The time evolution of the circularity at 2-hour intervals in MDA-MB231 and MCF10A spheroids is shown in Figs. 4b and 4c, respectively. In MDA-MB-231 spheroids, compression led to a significant change from circular shape to a less circular shape over time. The circularity values changed from 0.832 to 0.427 under compression and from 0.831 to 0.669 under the control condition (Fig. 4b). In contrast, MCF10A spheroids maintained a close-to-circular shape at all time points regardless of compression or control conditions. The circularity values ranged from 0.8995 to 0.859 under both control and compressed conditions (Fig. 4c). The observed significant decrease in circularity of MDA-MB-231 spheroids occurred after 6 hours of compression, suggesting a time-dependent response to compression.

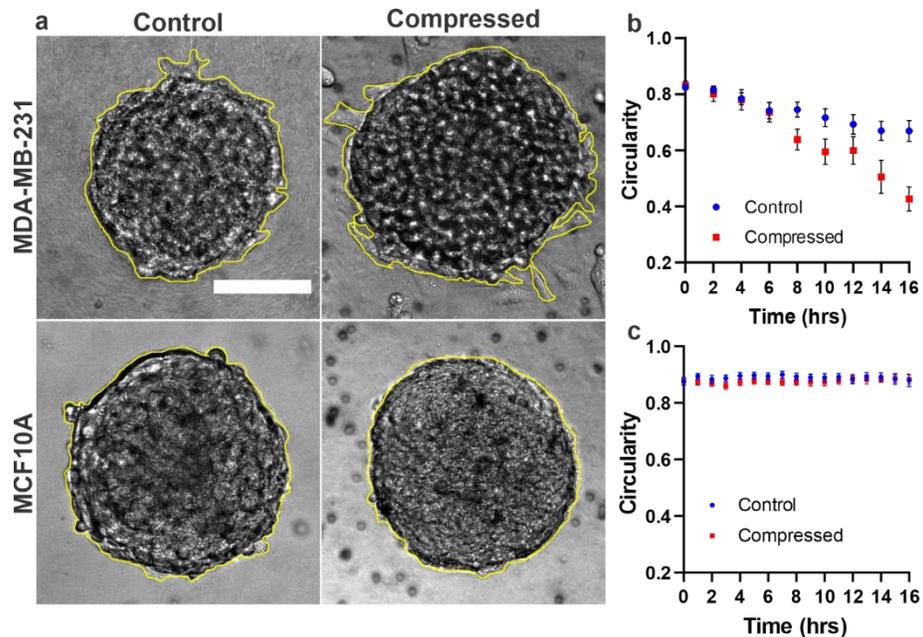

**Figure 4: Compression regulates morphology of MDA-MB-231 spheroids but has no apparent effect on MCF10A spheroid morphology. a)** Brightfield images of MDA-MB-231 and MCF10A spheroids embedded in 3.5 mg/mL collagen in control and compressed condition at t = 16 hours. The scale bar is 100 μm **b)** Change in circularity of MDA-MB-231 spheroids over a period of 16 hours in control and compressed conditions. **c)** Change in circularity of MCF10A spheroids over a period of 16 hours in control and compressed conditions. In (b) and (c), the spheroid outlines were generated manually and then the circularity of spheroids were calculated using ImageJ. Data in figure (4b) are from three separate experiments. A total of 17 spheroids were used in control and 13 in compressed condition. Data in figure (4c) are from three separate experiments. A total of 21 spheroids were used in control and 19 in compressed condition.

Here we used circularity as a measure of tumor spheroid peripheral roughness. A sphere corresponds to a geometry where the surface volume ratio is the smallest; hence, the amount of least roughness at the spheroid ECM interface. We found that MDA-MB-231 spheroids became rougher over time under compression compared to the control. Upon careful examination of the images, we found that the roughness was due to protrusions around the spheroids, a first step before cells detached from the spheroids. In parallel, the MCF10A spheroids showed no clear response to compression. We note that the roughness of tumor periphery has been used as a diagnostic tool for tumor malignant tumors in previous literature for melanoma[31].

*Compression promoted the invasion of MDA-MB-231 cells into the ECM but not MCF10A cells.*

Cells that broke away from the spheroids and invaded into the ECM were manually counted using time-lapse bright-field images (see Fig. 5a and Movie S1). The bright red lines highlight cells that invaded into the ECM. We found that on average, 7.23 ± 1.8 MDA-MB-231 cells per spheroid broke away from compressed tumor spheroids during the 16-hour observation time in contrast to 0.05 ± 0.01 cells per spheroid in control (see Fig. 5c). Similarly, we investigated the invasion of MCF10A cells into the ECM, but we did not observe any cells invading into the ECM under either compressed or control conditions during 16 hours of experiment (Fig. 5b and Movie S2).

Taken together, both circularity and single-cell invasion studies demonstrated that compression promoted MDA-MB-231 tumor cell invasion but had no apparent impact on MCF10A cell invasion. In addition, we observed time dependency in both the spheroid circularity and single-cell breakout experiments. In MDA-MB-231 spheroids, we observed significant invasion only after the first 6-8 hours (see Movie S1), and this was observed quantitatively in both circularity and single-cell invasion. MCF10A cells may take longer to invade into the ECM than the 16 hours that we followed them for; thus, longer experiments would provide more information on the invasion patterns of MCF10A spheroids. We note that some MCF10A spheroids compacted when embedded in ECM, however these spheroids were not considered for data analysis (see Fig. S3).

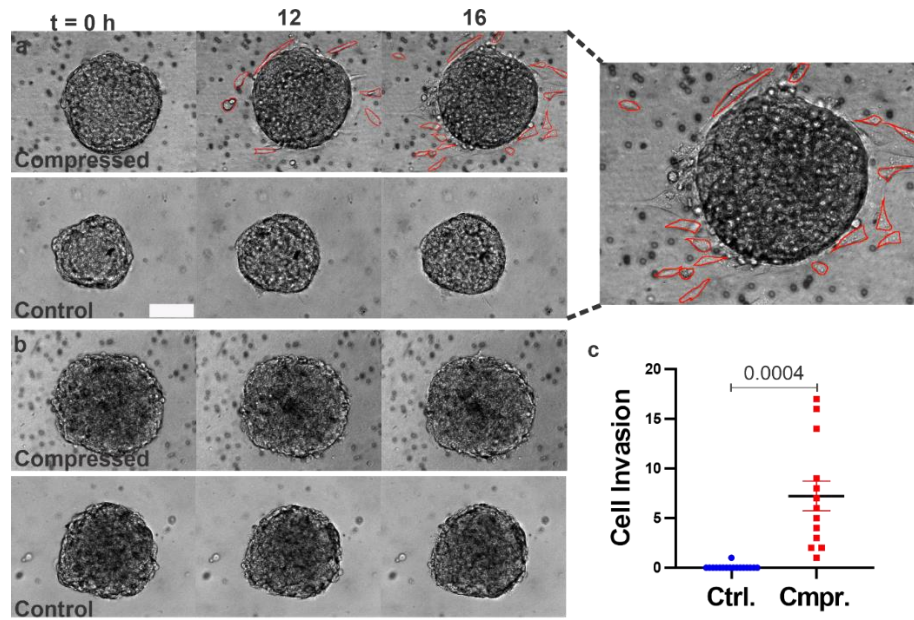

**Figure 5: Compression promoted invasion of MDA-MB-231 cells into the ECM and had no observable effect on MCF10A spheroids. a)** Time-lapse images of MDA-MB-231 spheroids in control and compressed conditions. On the right is the zoomed in image of an MDA-MB-231 tumor spheroid at t = 16 hour. Cells that invaded into the ECM are outlined with bright red lines. The scale bar is 100 µm. **b)** Time-lapse images of MCF10A spheroids in control and compressed conditions. **c)** Cells that broke away from the spheroid and invaded into the ECM were manually counted using ImageJ for MDA-MB-231 spheroids. A parametric Welch's t-test compared to the control group.

**Conclusion and Future Perspectives**

In this study, we developed a 3D in vitro tumor spheroid model to quantitatively investigate the effects of compression on tumor spheroid mechanics and invasion. Our results showed that compression had differential effects on the mechanics and invasion of malignant and non-tumorigenic epithelial spheroids. Upon compression, malignant MDA-MB-231 cells became more motile within the spheroids and more invasive into the ECM, whereas in non-tumorigenic MCF10A spheroids, motility within spheroids was suppressed, and no invasion was observed.

Differential responses to mechanical compression of malignant and non-tumorigenic cells can be caused by the mechanosensing ion channel, Piezo 1. It has been reported that malignant cells have substantially higher levels of Piezo1 than non-tumorigenic MCF10A cells[32][33]. Previous studies have also shown that Piezo1 activates integrin [34-36]. Considering the higher levels of Piezo1 in malignant cells like MCF7 and MDA-MB-231, compared to nontransformed cells like MCF10A, the lower levels of Piezo1 in normal epithelial cells could result in less activation of integrin. This, in turn, could lead to reduced invasion. Additionally, normal cells have lower levels of integrin compared to malignant cells[37], which further contributes to the observed differential response. Overall, our results are consistent with what was reported in 2D wound healing assays, in that malignant cancer cells are more mechanosensitive and compression selectively stimulates tumor cell invasion compared to normal cells [38].

The next step is to understand the molecular mechanisms underlying the differential responses of malignant and non-tumorigenic spheroids to compression. It is important to first identify the key molecular players using a general approach first, including RNA sequencing, to gain an unbiased insight into all transcripts. Systematic studies on breast cancer cells with different invasive potentials, including MCF7, T47D can reveal and validate the key molecular players. It would be interesting to conduct studies on other tumor types, such as brain tumors, in which the malignant state of the tumors is often highly compressed. Systematic investigation of the effects of compression on different tumors can potentially answer the question whether mechanosensitivity is correlated with tumor malignancy. Here, our work established a straightforward *in vitro* model to investigate the effects of compression on tumor spheroids, allowed for studies of their subsequent invasion in a 3D microenvironment, and highlighted the importance of compression in tumor mechanics and invasion.

**Materials and Methods:**

**Cells, spheroids, and 3D spheroid culture preparation.**

*Cells*. Metastatic breast adenocarcinoma cells (MDA-MB-231 cell line) and non-tumorigenic epithelial cells (MCF10A cell line) were provided by the Cornell Center of Microenvironment and Metastasis. MDA-MB-231 were cultured for up to 20 passages, and were used at 50-70% confluency[30]. The growth medium for MDA-MB-231 cells was composed of DMEM high glucose medium (Catalog No. [Cat.] 11965092, Gibco, Life Technologies Corporation, Grand Island, NY), 10% fetal bovine serum (Cat. S11150, Atlanta Biologicals, Lawrenceville, GA), and 1% antibiotics (100 units/mL penicillin and 100μg/mL streptomycin, Cat. 15140122, Gibco). MCF-10A cells were cultured up to passage 10 and used at 70–90% confluency. The growth medium for MCF-10A cells was composed of DMEM/F-12 medium (Cat. 11320033, Gibco), 5% donor horse serum (Cat. S12150, Atlanta Biologicals), 20ng/mL human EGF (Cat. PHG0311, Gibco),

0.5µg/mL hydrocortisone (Cat. H0888–1G, SigmaAldrich, St. Louis, MO), 100 ng/mL Cholera Toxin (resuspend at 1 mg/ml in sterile DI H2O, Cat. C8052-.5MG, Sigma-Aldrich), 10µg/mL insulin (Cat. 10516-5ML, Sigma-Aldrich), and 5% antibiotics (Gibco). We note that DMEM/F12 media was used for the preparation of both MDA-MB-231 and MCF10A spheroids. MDA-MB-231 expressing EGFP and MCF-10A cells expressing were kind gifts from Dr. Joseph Aslan at the Oregon Health & Science University. Fluorescently labeled MDA-MB-231 and MCF-10A cells were cultured in the same way as the non-labeled cells and were used for the co-culture spheroid experiments.

*Spheroids.* Uniformly sized spheroids were generated using a specially designed microwell array platform (see Fig. S1)[15, 23, 40]. Briefly, a silicon master of 18 X 18 microwell array was fabricated in the Cornell Nanofabrication Facility (CNF) using a one-layer photo-lithography method. Then the microwells were patterned on a 1 mm thick and 1x1 cm size agarose gel using a soft lithography method (see Fig. S1). Each microwell is cylindrical in shape with a diameter of 400 µm and depth of 400 µm. The agarose gel surface provides low adhesion of cells, thus promoting clustering of cells which results in spheroid formation. 6 individual microwell arrays ~ 1 cm in size were then placed in 6 wells of a 12- well plate (Cat. #: 07-200-82, Corning). Within each well of the 12-well plate, 3 million cells (1:20 ratio of fluorescently labeled: non-labeled cells) suspended in 2.5 ml of DMEM/F12 growth medium were introduced. The plate was then gently placed in 5% CO2 incubator at 100% humidity for four days. On day 3, DMEM/F12 spheroid growth medium was replenished. We note that the architecture of MDA-MB-231 spheroids is different from MCF10A spheroids. MCF10A spheroids are very compact and are formed overnight while MDA-MB-231 spheroids are less compact and take 3 to 5 days to form uniform sized spheroids. We harvested the spheroids on day 5 for both the cell lines. Tumor spheroids were collected from two arrays of microwells for each experiment and filtered by a Falcon Cell Strainer (Cat. #: 352360, Corning) with 100 µm pores to ensure the uniformity of the spheroid size. It is important to note that rich media and 5 days culture is important to form uniform sized MDA-MB-231 tumor spheroids.

*Spheroid embedded ECM.* To make 3D tumor spheroid cultures, we suspended spheroids in a 3.5 mg/ml type I collagen matrix (rat tail tendon Cat. #: 354249, Corning). Briefly, for each experiment, 200 µl of spheroid embedded collagen mixture was prepared with a collagen concentration of 3.5 mg/ml. To do this, 73.68 µl of collagen stock (9.5 mg/ml) was first titrated with 1.62 µl 1 N NaOH and 20 µl 10X M199 (Cat. #: M0650-100Ml, Sigma) to yield a final pH of ~7.4[41]. Then, 104.69 µl of spheroids with DMEM or DMEM/F12 GM for MDA-MB-231 and MCF10A spheroids respectively was added to reach a final volume of 200 µl. On average there were 1620 spheroids per ml of collagen. The final average spheroid concentration was approximately 4-5 spheroids per *in vitro* device (about 1 spheroid per mm$^2$ under the top view).

*In vitro* **device setup.** A silicon master with 7 cylindrically shaped pistons of 4 mm diameter and 115 µm depth was fabricated in the Cornell Nanofabrication Facility (CNF) using one-layer photolithography. 1 mm thick PDMS membrane with 7 wells were patterned from the silicon master using soft lithography and was placed at the base of one of the wells in a 24-well plate. The 1 mm thickness of the PDMS membrane is critical to ensure that Transwell insert sits perfectly on top of the PDMS well. Spheroid-embedded collagen was introduced in the PDMS well. A Transwell insert was placed on top of the PDMS well after the collagen was polymerized (see Fig. 1). A static load of 35 grams (metallic ring) was then placed on top of the Transwell insert to compress the spheroids. For sterility, PDMS wells were autoclaved and both the PDMS wells and the 24-well plate (Cat #: 353047) were treated with oxygen plasma (Harrick Plasma Cleaner PDC-001, Harrick Plasma, Ithaca, NY) for 1 minute on high power mode.

*Surface activation.* To ensure optimal surface properties for the binding of the collagen, the surface of the PDMS well was activated using 1% Poly(ethyleneimine) (Cat. P3143-100ML, Sigma-Aldrich) for 10 minutes followed by thorough rinsing, using sterilized DI water, and treating the device with 0.1 % Glutaraldehyde (Cat. 16019, Electron Microscopy Sciences, Hatfeld, PA) for 30 minutes. The PDMS wells were again thoroughly rinsed three times. The PDMS wells were filled with sterilized DI water and left in the biohood overnight at room temperature.

*3D spheroid seeding.* On the day of an experiment, the PDMS wells-bonded 24 well-plate was placed on an ice pack for the entire device setting up time. After allowing the PDMS wells to cool, 2.5 µl of spheroid-embedded collagen solution was added to each PDMS well. The well-plate was then placed on a larger petri dish padded with wet tissues, and the petri dish was then placed in the incubator to allow collagen polymerization at 37 C and 5% CO2 for 30 minutes. We note that the temperature ramping rate during polymerization plays an important role in collagen structure[42]; to get a small and homogenous pore size network we used fast warming in our setup. To achieve fast warming the petri dish was placed on a metallic block (which was at 37 C) instead of directly placing the petri dish in the incubator. Slow warming results in an inhomogeneous large pore size network[15]. Following the polymerization, 1 ml of DMEM or DMEM/F12 growth medium was added for MDA-MB-231 or MCF10A experiment setup respectively. The well plate was then transferred to the microscope stage enclosed by an environmental control chamber (WeatherStation, PrecisionControl LLC), which was kept at 37 °C, 5% CO2 and about 70% humidity. Here t=0 is the time at which the imaging started which is approximately 2 hours after the spheroid embedded collagen was polymerized. Each experiment had 2 compressed conditions where a load was applied using Transwell insert (Cat #: 353097, Corning) and 2-3 control wells where no static load was applied on the spheroids. Each experiment was repeated 3 times.

**Imaging and data analysis:**

All images were taken using an inverted epi-fluorescent microscope (IX81, Olympus America, Center Valley, PA, USA) with a CCD camera (ORCA-R2, Hamamatsu Photonics, Bridgewater, NJ, USA). In all the experiments the middle z-plane of the spheroids was captured using a 20X objective (Olympus, NA = 1) in bright field and in green fluorescence. The light source for fluorescence imaging was provided by the X-Cite series 120PC unit (Excelitas Technologies, Waltham, MA, USA). The scope has a stage incubator (Precision Plastics Inc., Beltsville, MD, USA) that maintained a temperature of 37 °C, humidity of ~70%, and 5 % CO2 level. The setup was placed on the automated X-Y microscope stage (MS-2000, Applied Scientific Instrumentation, Eugene, OR), and images were taken every 10 minutes for 16 hours using CellSens software (Olympus America, Center Valley, PA, USA). For each experiment brightfield and GFP images were taken.

Using the brightfield time-lapse images, the circularity of spheroids was calculated by manually drawing the outlines of the spheroids and then applying the circularity measurement parameter in ImageJ. Circularity is defined as ($4\pi \, Area/perimeter Perimeter^2$) [43]. The cell trajectories were obtained using the manual tracker in ImageJ (National Institute of Health) using the GFP time-lapse images. Single cell migration parameters of speed, persistence length and mean squared displacement were calculated using these trajectories[30]. The cell speed is defined as the total length of the track divided by the time duration. The cell persistence length is defined as the distance between the cell starting and ending positions divided by the length of the cell trajectory. Parametric t-test (Welch test) was carried out using Prism (GraphPad Software,Inc., La Jolla, CA).

**Supplementary Figures:**

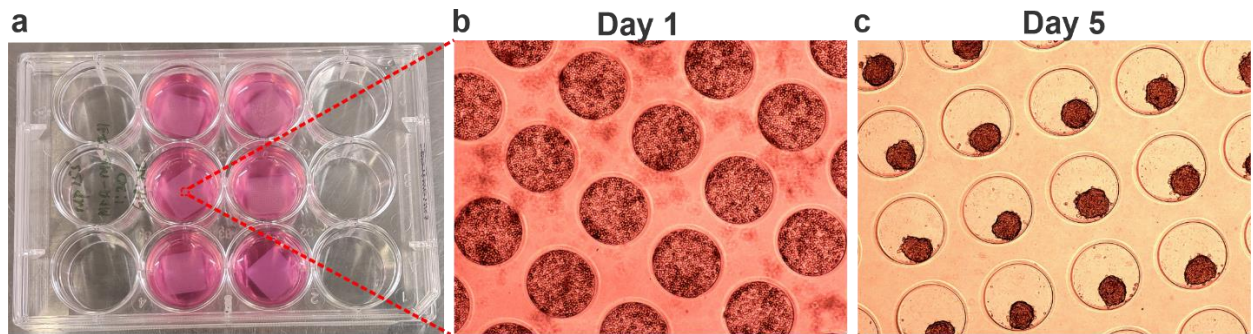

**Figure S1: High throughput tumor spheroid formation assay.** A) Image of a 12 well plate containing high throughput agarose based microwell array for spheroid formation. Each microwell device has 18x18 microwells patterned agarose gel membrane, and each microwell is 400 µm in diameter and depth. In a typical experiment 2 of these arrays are used to harvest spheroids. **B)** Micrograph of tumor spheroid formation in a microwell array. A total of 3 million MDA-MB-231 cells are seeded on day 1 to each microwell array. **C)** The prepared spheroids are harvested on day 5 and the media is changed regularly every 2-3 days. MCF10A spheroids were made in the same device.

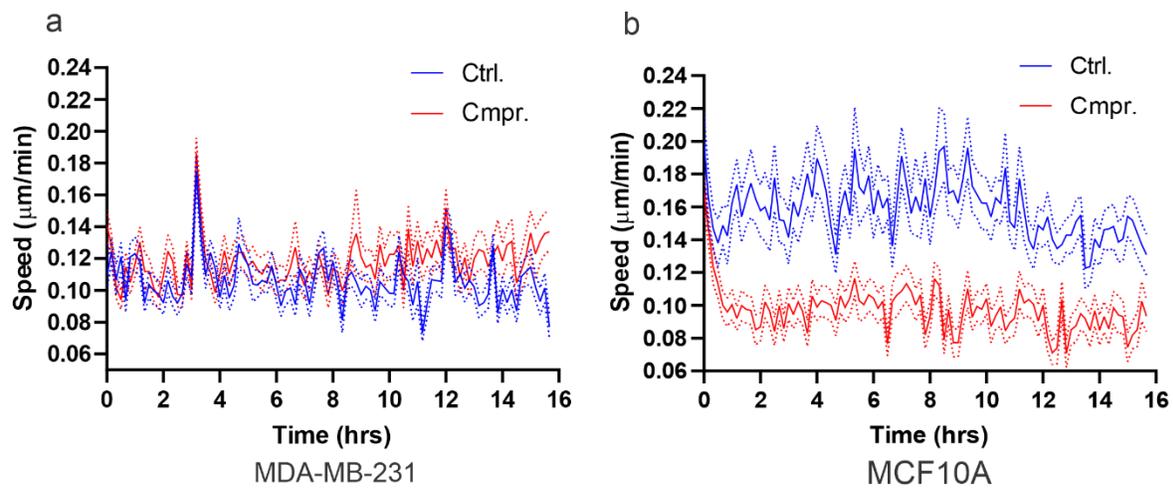

**Figure S2: Time evolution of speed in MDA-MB-231 and MCF10A spheroids. A)** Speed of MDA-MB-231 cells within spheroids **B)** Speed of MCF10A cells within spheroids as a function of time under compressed and control conditions.

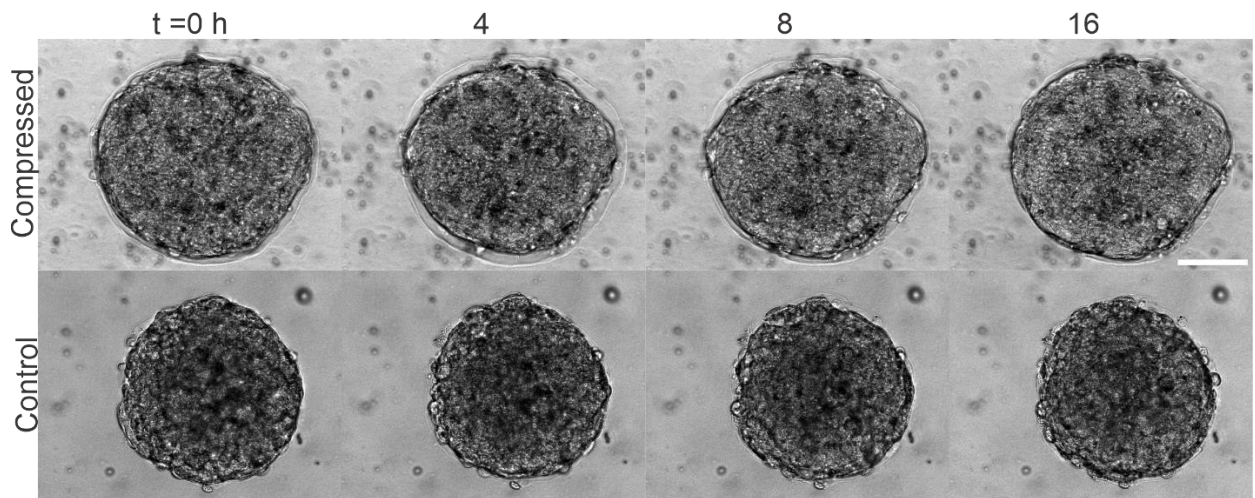

**Figure S3: Compaction of MCF10A spheroids in 3.5 mg/ml collagen.** Micrographs of MCF10A spheroids embedded in 3.5 mg/ml collagen taken at t = 0 hours for compressed and control condition. Scale bar is 50 μm.

## Acknowledgements:

This work was supported by a grant from the National Institute of Health (Grant No. 5R01CA221346-06). MP and Wu thanks the helpful discussions with Jan Lammerding, Claudia Fischbach, Jennifer Schwarz, and M. Lisa Manning. This work was performed in part at the Cornell NanoScale Facility, a member of the National Nanotechnology Coordinated Infrastructure (NNCI), which is supported by the National Science Foundation (Grant NNCI-2025233). JES is the Betty and Sheldon Feinberg Senior Faculty Scholar in Cancer Research.

## Author contributions:

M.P., J.E.S. and M.W. designed the research, contributed analytic tools, analyzed data and wrote the paper. M.P. performed the experiments. Y.J.S. fabricated the microwell device. M.K. and H.J.D. analyzed data.

## Data availability:

All the raw data and analyzing tools for this study are available upon request to the corresponding author.

## Competing interests:

The authors declare no competing interests.